\documentclass[twocolumn,showpacs,showkeys,preprintnumbers,amsmath,amssymb,floatfix,nofootinbib]{revtex4}

\usepackage{graphicx}
\usepackage{bm}
\usepackage{url}
\usepackage{xcolor}
\def\be{\begin{equation}}
\def\ee{\end{equation}}

\begin{document}

\title{Self-consistent Skyrme QRPA for use in axially-symmetric nuclei of
arbitrary mass}

\author{J.\ Terasaki}
\affiliation{Department of Physics and Astronomy, University of North
Carolina, Chapel Hill, NC 27599-3255}
\author{J.\ Engel}
\affiliation{Department of Physics and Astronomy, University of North
Carolina, Chapel Hill, NC 27599-3255}

\begin{abstract}
We describe a new implementation of the quasiparticle random phase
approximation (QRPA) in axially-symmetric deformed nuclei with Skyrme and
volume-pairing energy-density functionals.  After using a variety of tests to
demonstrate the accuracy of the code in $^{24,26}$Mg and $^{16}$O, we report
the first fully self-consistent application of the Skyrme QRPA to a heavy
deformed nucleus, calculating strength distributions for several $K^\pi$ in
$^{172}$Yb.  We present energy-weighted sums, properties of
$\gamma$-vibrational and low-energy $K^\pi$=0$^+$ states, and the complete
isovector $E$1 strength function.
The QRPA calculation reproduces the properties of the low-lying $2^+$ states as
well or better than it typically does in spherical nuclei.
\end{abstract}

\pacs{21.60.Jz}
\keywords{QRPA, deformed, strength function}
\maketitle

\section{Introduction}

The quasiparticle random phase approximation (QRPA) \cite{Rin80,Bla86} has a
long history in nuclear physics.  Its virtues include applicability to many
types of excitation across the isotopic chart, preservation of energy-weighted
sum rules, and elimination of spurious motion.  In addition, the QRPA has
several appealing interpretations; it is both a boson approximation for
collective modes and the small-amplitude limit of the time-dependent
Hartree-Fock-Bogoliubov (HFB) approximation.  Its downside, traditionally, has
been a limited ability to describe large-amplitude motion and complicated
non-collective states, deficiencies that prompted the development of several
more complicated methods, as, e.g., in Refs.\ \cite{Tso04,Sar04}.

Recent years, however, have seen a revival of the QRPA, despite its drawbacks.
The primary reason is the increasing connection between nuclear mean-field
theory and density-functional theory (DFT) \cite{Pet91,Fio03}.  The notion that
Hartree-Fock or HFB calculations can be relevant beyond their naive range of
validity has motivated attempts to describe a wide-range of nuclear properties
in mean-field theory and extensions.  The QRPA is the most straightforward
extension that fits into the DFT paradigm; to the extent that the energy
functional used in HFB calculations is exact, the QRPA provides the exact
linear (i.e.\ small-amplitude) response function in the adiabatic limit
\cite{Fio03}.  Combined with its other features, its connection with DFT makes
the QRPA an important tool in attempts both to develop a ``Universal Nuclear
Energy-Density Functional'' (UNEDF), and to apply the functional to, e.g.\
nuclear astrophysics \cite{Arn03}. 

The prototype energy-density functional is of Skyrme form, corresponding
roughly to effective interactions that have zero range, with derivatives
simulating finite-range effects.  In the last five or ten year, a number of
groups have developed self-consistent (Q)RPA codes for use with these
functionals or similar finite-range and relativistic versions, first in
spherical nuclei, (see Ref.\ \cite{Ter05} and references therein), and then in
axially symmetric deformed nuclei \cite{Yam04,Pen08,Per08,Yos08,Los10}.  Heavy
deformed nuclei are still problematic, however.  Though the deformed RPA,
without pairing, is now tractable in heavy nuclei \cite{Ina09}, a separable
approximation to the Skyrme-QRPA equations has been applied in such nuclei
\cite{Nes06,Sev08}, and efficient new methods for solving the full QRPA
equations are promising \cite{Ina09,Toi10}, the numerical complexity of
deformed systems has so far limited fully self-consistent Skyrme-QRPA
calculations to nuclei with $A \lesssim 40$.  In this paper, we present a
highly parallelized version of the Skyrme QRPA that we are beginning to apply
to heavy deformed nuclei.  After discussing the structure of the code and
demonstrating its accuracy, we present a preliminary application to the nucleus
$^{172}$Yb.  We focus on results here, postponing most of the formalism to a
later publication.

Our QRPA code comes in several versions, developed successively as we
progressed from tests in light nuclei to full-fledged calculations in heavy
nuclei.  All versions treat the entire Skyrme + Coulomb functional in a way
that is completely consistent with HFB calculations (restricted for the time
being to ``volume'' pairing).  In addition, all the versions preserve axial and
parity symmetries and the time-reversal invariance of the ground state, and
therefore require an HFB code that produces single-quasiparticle wave functions
of the two variables $r\equiv \sqrt{x^2+y^2}$ and $z$, with
$M\equiv\langle{j_z}\rangle$ a good quantum number.  Finally, all diagonalize
the traditional QRPA $A$, $B$ matrix \cite{Rin80} in the ``$M$-scheme'', and
use the rigid-rotor approximation \cite{Boh75,Sol89}, with the deformed
QRPA solutions as intrinsic states, to calculate transition strength.  The
versions differ, however, in the basis in which the HFB equations are solved,
in the basis in which the QRPA matrix is constructed, and in the way wave
functions are represented numerically.  Version a) takes quasiparticle-basis
wave functions from the transformed-oscillator code HFBTHO \cite{Sto05} (though
we use the ordinary harmonic-oscillator basis), represents the wave functions
on an equidistant mesh, and constructs the QRPA matrix in the quasiparticle
representation.  Version b) substitutes the Vanderbilt ``cylindrical-box''
$B$-spline-based HFB code \cite{Bla05} for HFBTHO.  Version c) modifies the
QRPA part of version b) by using the canonical-quasiparticle basis,
represented with $B$-splines, in place of the quasiparticle basis to speed
calculation and save memory.  The vast majority of the computing time in all
versions is in the construction of the QRPA matrix, each element of which
requires a series of two-dimensional integrals for the Skyrme interaction, and
an additional multipole expansion for the Coulomb interaction.  The set of
matrix elements can be divided among many thousands of processors so that the
calculation is manageable on fast supercomputers.

Section II below describes tests in relatively light systems.  Section III
presents an application to heavy nuclei.

\section{Tests}

To display the accuracy of our codes, we show the results of several tests in
nuclei with $A<40$.  We start with spurious states.  A fully self-consistent
and numerically perfect QRPA will completely separate spurious states from
physical ones, and put them at zero energy.  Small numerical errors can spoil
the treatment of spurious states, however, so any calculation that
separates them well has passed a serious test.

\begin{figure}
\includegraphics[width=8cm]{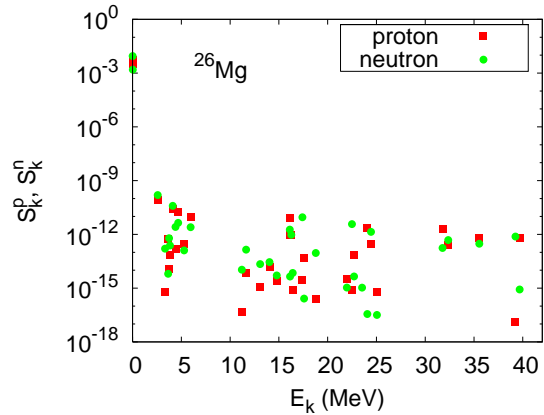}%
\caption{\label{fig_nstr} (Color online) Transition strength for the proton and
neutron number operators to $K^\pi=0^+$ states in $^{26}$Mg with the Skyrme
functional SKP (see text).}
\end{figure}

Figure \ref{fig_nstr} shows transition matrix elements of the number operators,
\begin{equation}
S^\tau_k = | \langle 0 | \hat{N}_\tau | k \rangle |^2\,,
\label{particle_number_strength}
\end{equation}
with $k$ labeling excitations and $\tau$ protons/neutrons, to states with
$K^\pi=0^+$ in a calculation of $^{26}$Mg.  To obtain the QRPA wave functions,
we use version a) of our code with the Skyrme interaction SkP
\cite{Dob84}, restricting ourselves to three harmonic oscillator shells.  The
HFB calculation yields pairing gaps of $\Delta_p=1.681$ MeV (protons),
$\Delta_n=1.426$ MeV (neutrons) and no quadrupole deformation in this small
single-particle space.  This result does not imply that a calculation in a
larger space also gives $\beta=0$ \cite{Sto03}.  We use a small space to allow
a complete QRPA treatment of two-quasiparticle excitations; a full
separation of the spurious $J^\pi=0^+$ strength associated with the
particle-number violation requires no less.  And we indeed achieve essentially
perfect separation. The figure shows negligible strength to all excitations
except the two spurious states, which come out at $E=0.008$ MeV and 0.024 MeV.

\begin{figure}
\includegraphics[width=8cm]{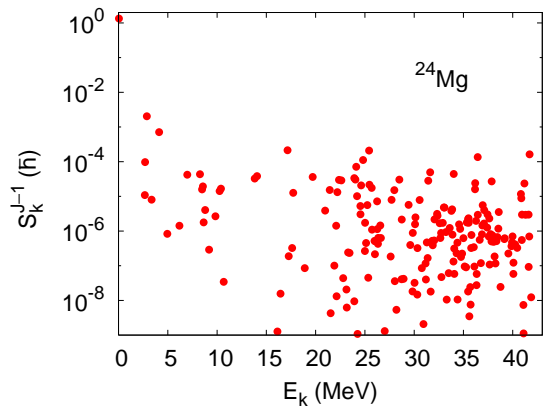}%
\caption{\label{fig_jm1str} (Color online) Transition strength for the operator
$J_{-1}$ to $K^\pi=1^+$ states in $^{24}$Mg with SkP.}
\end{figure}

Next, we turn to spurious rotation, arising in the $K^\pi$ = 1$^+$ channel.
Here the nucleus must be deformed.  Our HFB solution, again with SkP but for
$^{24}$Mg in five oscillator shells, yields $\beta$ = 0.28, $\Delta_p$ = 0.034
MeV, and $\Delta_n$ = 0.131 MeV.  Figure \ref{fig_jm1str} displays the
resulting transition strengths for the operator $J_{-1} \equiv J_x-iJ_y$.  Most
of the strengths are four order of magnitude smaller than that of the spurious
state at $E$ = 0.045 MeV.  

\begin{figure}
\includegraphics[width=8cm]{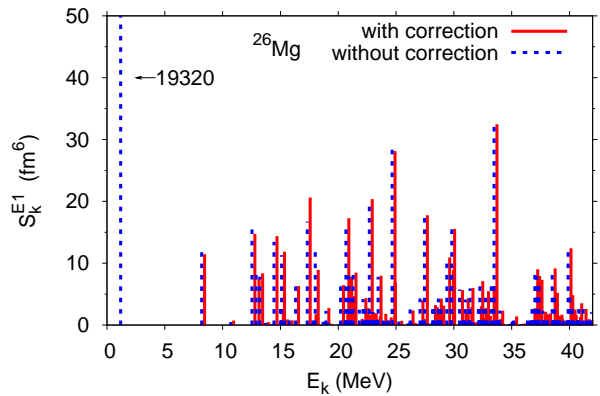}%
\caption{\label{fig_e1k0-str} (Color online) IS $E1$ transition strength to
$K^\pi=0^-$ states in $^{26}$Mg with SLy4. 
Solid lines represent strength
that is corrected to eliminate residual spurious contributions.}
\end{figure}

Finally, we examine spurious translational motion.  Here, an oscillator basis
for HFB is not optimal, and so we employ versions b) and c) of our code (they
give identical strength functions), which use wave functions in a
large cylindrical box.  For these calculations, in $^{26}$Mg, we choose the box
to have size $r_\textrm{max}$ = $z_\textrm{max}$ = 10 fm and use the Skyrme
functional SLy4 \cite{Cha98}.  We take a pairing strength $V$ = $-$140.348 MeV
fm$^3$ and include quasiparticle states with energy $\leq$ 300 MeV in the
density and pairing tensor.  These parameters yield $\beta$ = $-$0.27,
$\Delta_p$ = 1.365 MeV, and $\Delta_n$ = 0.002 MeV.  Figure \ref{fig_e1k0-str}
shows isoscalar (IS) $E$1 transition strengths, for which the excitation
operator is $\sum_{i=1}^A r_i^3\, Y_{1\mu}(\Omega_i)$, to states
with $K^\pi=0^-$.  The figure contains two sets of lines, the second of which
adds a correction term to the IS operator (via the prescription of
Ref.\ \cite{Yos08}) to remove residual spurious strength from physical
excitations.  The difference between the two sets is very small in all the
physical states shown, and completely negligible in higher-energy
states.  In sum, our code is accurate enough to handle spurious states in this
mass region extremely well.

\begin{table}
\caption{\label{tbl_e_str_JM} Comparison of energies and (corrected) IS $E1$
strengths for the three lowest-lying $J^\pi=1^-$ states in $^{16}$O,
calculated with spherical ($J$-scheme) code and the current ($M$-scheme) code
and SLy4.  The correction barely changes the strength for the two physical
excited states, but reduces that of the spurious (lowest) state by many orders
of magnitude.}
\begin{ruledtabular}
\begin{tabular}{rlrl}
\multicolumn{2}{c}{$J$-scheme, $J^\pi=1^-$} & \multicolumn{2}{c}{$M$-scheme,
$K^\pi=0^-$} \\ \hline
   $E$\textrm{\ \ } & \ \ $S^{\textrm{IS}E1}_{\textrm{cor}}$ & $E$\textrm{\ \ } & \ \ $S^{\textrm{IS}E1}_{\textrm{cor}}$ \\
   (MeV) & \ \ (fm$^6$)                & (MeV) & \ \ (fm$^6$)                \\
\hline
0.323 & 7.051$\times 10^{-5}$ &  0.472    & 1.298$\times 10^{-4}$           \\
7.500 & 1.461$\times 10$     &  7.440    & 1.433$\times 10$               \\
10.610 & 5.739$\times 10^{-2}$ & 10.681 & 4.283$\times 10^{-2}$ 
\end{tabular}
\end{ruledtabular}
\end{table}

\begin{figure}[t]
\includegraphics[width=8cm]{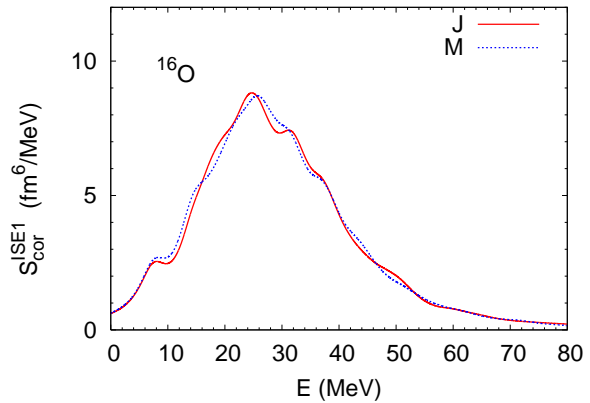}%
\caption{\label{fig_strfn_ise1_o16} (Color online) Full IS $E1$ strength
function corresponding to Tab.~\ref{tbl_e_str_JM}. The letters \textsf{J} and
(\textsf{M}) denotes the $J$- ($M$-) scheme calculations.}
\end{figure}

Since we have done extensive calculations with a spherical $J$-scheme code over
the past few years \cite{Ter06}, we can test our current codes further by
comparing their results with those obtained in the $J$ scheme.  Table
\ref{tbl_e_str_JM} shows energies and IS $E$1 transition strengths (with the
correction mentioned above) for the three lowest $J^\pi=1^-$ levels, along
with the corresponding $K^\pi$ = 0$^-$ energies and strengths from version b)
of our current code, in the spherical nucleus $^{16}$O.  The two codes take
wave functions from entirely different HFB codes:  a slightly modified version
of HFBRAD \cite{Dob84} called HFBMARIO for the spherical QRPA and the
Vanderbilt HFB code for the deformed QRPA.  The first state on each side
of the table is the spurious state, with very small strength because of the
correction.  The next two, both genuine excitations, are nearly the same in
both energy and strength in the two calculations.  The full strength function,
folded with a Lorentzian of width 3 MeV (see in Eq.\ (1) in Ref.\ \cite{Ter05})
displayed in Fig.\ \ref{fig_strfn_ise1_o16}, shows the same level of
agreement.  The very small differences in the continuum are due to differing
box boundary conditions: the spherical calculation is in a spherical box with
radius 20 fm and the deformed calculation is in the same cylindrical box
we used for $^{26}$Mg.  

\begin{figure}[b]
\includegraphics[width=8cm]{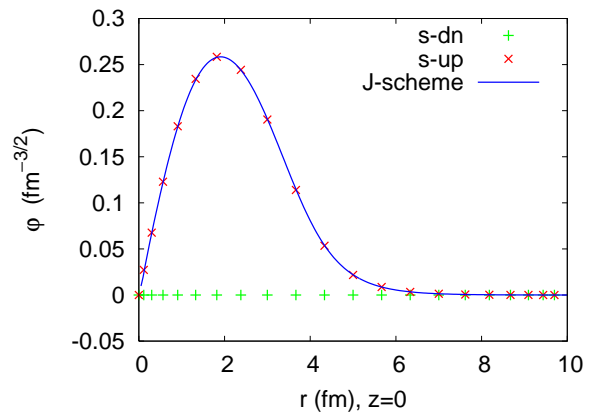}%
\caption{\label{fig_can2311s} (Color online) Comparison of proton 0$p_{3/2}$
canonical-basis wave function produced by $J$- and $M$-scheme codes for 
$^{22}$O at $z=0$.  The labels \textsf{s-dn} and \textsf{s-up} denote the
spin-down and spin-up components in the $M$-scheme calculation.  In the $J$
scheme, the spin-down component is identically zero.}
\end{figure}

Before moving to heavy nuclei, we display the results of one more test --- this
time simultaneously checking the Vanderbilt HFB code \cite{Bla05} underlying
our deformed QRPA, and our procedure for constructing the canonical basis by
diagonalizing the density matrix (which is the same as in Ref.\ \cite{Ter05}).
Figure \ref{fig_can2311s} displays the proton $0p_{3/2}$ canonical basis wave
function in $^{22}$O produced by both the spherical and deformed procedures.
The agreement is perfect and far from trivial.  Though $\Delta_p =0$ in this
calculation, we still construct the density matrix from the Vanderbilt-HFB
output and diagonalize it to obtain the deformed canonical wave function.
(There is no arbitrariness in this wave function because only one proton
$p_{3/2}$ state is occupied in Oxygen.)  The other bound-state wave functions
produced by the two procedures, though we don't display them, agree equally
well.  

\section{Heavy nuclei}

\begin{table}
\caption{\label{tbl_ewsr_172yb} Energy-weighted sums for strength functions, in
our QRPA calculations and from analytical sum-rule expressions.  We include
QRPA states with up to 90 MeV of excitation energy.  The units of the IS $E$1
sum are MeV fm$^6$, and those of the isovector (IV) $E$1 sum are MeV fm$^2$.
Those of all $E$2 sums are MeV fm$^4$.  The IS $E1$ strength has been
corrected to remove spurious components.
The contribution from negative values of $K$ is not included.
}
\begin{ruledtabular}
\begin{tabular}{llll}
   Transition & $K^\pi$     & \ \ QRPA             & \ \ Analytical        \\
   operator   & of solution & \ \                  &                       \\
\hline
  IS\ $E$1    & 1$^-$       & 1043560              & 1042413               \\
  IV\ $E$1    & 1$^-$       & 289.819              & 285.764               \\
  IS\ $E$1    & 0$^-$       & 2015266              & 2019465               \\
  IV\ $E$1    & 0$^-$       & 291.859              & 285.764               \\
  IS\ $E$2    & 2$^+$       & 64700                & 63877                 \\
  IV\ $E$2    & 2$^+$       & 20284                & 20076                 \\
  IS\ $E$2    & 1$^+$       & 76159                & 88197                 \\
  IV\ $E$2    & 1$^+$       & 28517                & 28174                 \\
  IS\ $E$2    & 0$^+$       & 97886                & 96867                 \\
  IV\ $E$2 & 0$^+$ & 31271 & 30874
\end{tabular}
\end{ruledtabular}
\end{table}

Having thoroughly tested several versions of the deformed QRPA, we apply
version c), representing the best combination of speed and accuracy, to the
nucleus $^{172}$Yb.  We set up the HFB calculation as follows: we use a 
``box'' with $r_\textrm{max}$ = $z_\textrm{max}$ = 20 fm,
cut off the quasiparticle spectrum at 60 MeV and take the maximum $z$-component
of quasiparticle angular momentum to be 19/2; these parameters define a
single-quasiparticle space with 4648 proton states and 5348 neutron states.  We
use the Skyrme functional SkM$^\ast$ \cite{Bar82} with volume pairing strengths
$V_p$ = $-$218.521 MeV fm$^3$ and $V_n$ = $-$176.364 MeV fm$^3$ (determined from
measured odd-even mass differences).  The calculation yields $\Delta_p$ = 1.248
MeV, $\Delta_n$ = 0.773 MeV, and $\beta$ = 0.34.

We input half the canonical wave functions, those with $j_z > 0$, in the QRPA,
and construct and truncate two-canonical-quasiparticle configurations in the
following way:  We define critical particle-particle and particle-hole
occupation probabilities $(v_\textrm{cut}^\textrm{pp})^2$ = 10$^{-6}$ and
$(v_\textrm{cut}^\textrm{ph})^2$ = 10$^{-10}$.  If both canonical states have
occupation probabilities $v_i^2$ and $v_j^2$ such that $v_i^2$, $v_j^2 > 1 -
(v_\textrm{cut}^\textrm{pp})^2$ or $v_i^2$, $v_j^2 <
(v_\textrm{cut}^\textrm{pp})^2$ , we omit the configuration.  We also omit
configurations with $v_i^2 < v_j^2$ for which $v_i^2/v_j^2 <
(v_\textrm{cut}^\textrm{ph})^2$. These cuts result in a QRPA matrix whose size,
while depending on multipolarity, is typically about 160,000 by
160,000.

Table \ref{tbl_ewsr_172yb} shows energy-weighted sums, alongside values
obtained from sum rules, for IS and isovector (IV) electric operators in all
$^{172}$Yb channels that we calculate.  The differences between the QRPA sums
and the sum rules are less than about 2\% except in the IS $K^\pi$ = 1$^+$
channel, where we were unable to adequately separate the spurious rotational
state without going to a larger space.  In other channels the spurious mode is
under better control.  In the IS $0^-$ and $1^-$ channels, we have some
contamination at low energies, but it is weak enough that the subtraction
procedure of Ref.\ \cite{Yos08} restores the sum rule nearly exactly.  In the
$0^+$ channel, as the table shows, the separation is quite good even without
subtraction.  The $1^+$ channel can be corrected as well, but doing so requires
a numerical procedure that we have not yet implemented. 

\begin{table}[t]
\caption{\label{tbl_be2} Energies and $B(E2;0^+\rightarrow 2^+)$'s for the
$\gamma$-vibrational and ``$\beta$- vib" states of $^{172}$Yb.  Experimental
data are from Ref.~\cite{Fah92} (see also \cite{natl_nucl_data}).  For
the definition of $B(E2)$, see, e.g., Ref.~\cite{Boh69}. }
\begin{ruledtabular}
\begin{tabular}{cccc}
                   &                       & Exp.           &   Cal.   \\
\hline
{$\gamma$-vib.}    &  $E$ (MeV)            & 1.466          &   2.261  \\
                   &  $B(E2)$ ($e^2$b$^2$) & 0.0433 +5$-$15 &   0.041  \\
\hline
``{$\beta$-vib.}'' &  $E$ (MeV)            & 1.117          &   1.390   \\
                   & $B(E2)$ ($e^2$b$^2$)  & 0.0081 17      & 0.00495 
\end{tabular}
\end{ruledtabular}
\end{table}

The accuracy of the energy-weighted sums in the $K^\pi=0^+$ and $2^+$ channels
indicates that our approach is reliable for low-lying quadrupole shape
vibrations.  Table \ref{tbl_be2} shows the energies and $B(E2;0^+\rightarrow
2^+)$'s of both the $\gamma$-vibrational $K^\pi = 2^+$ state and a low-energy
$K^\pi$ = 0$^+$ state with a significant $B(E2)$ that we denote by
``$\beta$-vib".  The quotation marks indicate that, unlike the clearcut
$\gamma$-vibrational state, the $0^+$ state has a somewhat smaller $B(E2)$ than
is typical of vibrational modes.  Both states have been studied experimentally,
e.g.\ in Ref.\ \cite{Fah92}. 

The agreement of both the energies of these states and their transition
strengths with measured values is at a level that is typical of QRPA
calculations in spherical nuclei.  In Ref.\ \cite{Ter08} we investigated a
large set of such nuclei, characterizing the quality of the QRPA by two
quantities: 
\begin{align}
 R_E &\equiv \ln ( E_\textrm{cal}/E_\textrm{exp})\,,\\
   R_Q &\equiv \ln \sqrt{ B(E2)_\textrm{cal}/B(E2)_\textrm{exp} }\,,\nonumber
\end{align}
where suffices cal and exp denote calculated and experimental.  The results in
Tab.\ \ref{tbl_be2} correspond to $R_E$ = 0.43 and $R_Q$ = $-$0.03 for the
$\gamma$-vibration, and $R_E$ = 0.22 and $R_Q$ = $-$0.25 for ``$\beta$-vib."
The histograms in Figs.\ 4 and 9 of Ref.~\cite{Ter08} show these values to be
near the most common values in spherical nuclei.

\begin{figure}
\includegraphics[width=8cm]{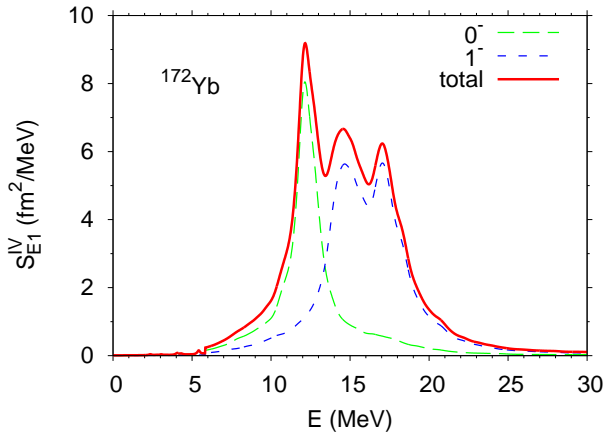}%
\caption{\label{fig_strfn_di_iv} (Color online) Predicted IV $E1$ strength
function in $^{172}$Yb,
with a folding width of 0.5 MeV (0.1 MeV for discrete states).
$0^-$ and $1^-$ indicate $K^\pi$ components 
(the curve labeled $K^\pi=1^-$ includes the contribution of $K^\pi=-1^-$).
}
\end{figure}

Figure \ref{fig_strfn_di_iv} displays the IV $E1$ strength function.
The thick curve is the sum of strengths in all channels, and can be compared
with experimental data.  The peaks of the $K^\pi$ = 0$^-$ and 1$^-$
distributions in Fig.~\ref{fig_strfn_di_iv} lie at different energies, as is
often the case in deformed nuclei. 
Though an experimental group reports a pygmy resonance at 3 $-$ 4 MeV
\cite{Voi01}, we see no indication of one in our calculation.  It has been
suggested that in spherical Sn isotopes such resonances involve configurations
beyond the natural ambit of the QRPA \cite{Tso04,Sar04}, but the issue is
unresolved, and in deformed nuclei has not been systematically investigated. 

In summary, we have developed a box-based Skyrme-QRPA code for
axially-symmetric even-even nuclei aimed at calculations throughout the
isotopic chart.  We rigorously tested the accuracy of the code in Mg isotopes,
paying particular attention to spurious states.  Though the code does not do
quite so well with spurious states in heavier nuclei under the restrictions on
space size that we currently impose, the agreement of energy weighted sums with
sum rules indicates that a) in channels with no spurious modes (e.g.\
$K^\pi=2^+$) our code is quite accurate, b) in the $K^\pi=0^+$ channel,
spurious admixtures are negligible, and c) in the IS $K^\pi$=$0^-$, and $1^-$
channels (and the $1^+$ channel in the future), admixtures are larger
but can be effectively removed.  In addition, calculations with larger
spaces should be possible.
Our immediate plans are to systematically
investigate the ability of modern density functionals, together with the QRPA,
to describe $\beta$ and $\gamma$ vibrations in the rare-earth nuclei.


\begin{acknowledgments}
This work was supported by the UNEDF SciDAC Collaboration under DOE grant
DE-FC02-07ER41457. We are indebted to Profs.\ V.\ E.~Oberacker and A.\ S.\ Umar
for giving us their HFB code and for technical support.  We used computers at
the National Energy Research Scientific Computing Center, the National
Institute for Computational Sciences at the University of Tennessee (TeraGrid),
the National Center for Computational Sciences, and the University of North
Carolina at Chapel Hill.  
\end{acknowledgments}


\begin{thebibliography}{31}
\expandafter\ifx\csname natexlab\endcsname\relax\def\natexlab#1{#1}\fi
\expandafter\ifx\csname bibnamefont\endcsname\relax
  \def\bibnamefont#1{#1}\fi
\expandafter\ifx\csname bibfnamefont\endcsname\relax
  \def\bibfnamefont#1{#1}\fi
\expandafter\ifx\csname citenamefont\endcsname\relax
  \def\citenamefont#1{#1}\fi
\expandafter\ifx\csname url\endcsname\relax
  \def\url#1{\texttt{#1}}\fi
\expandafter\ifx\csname urlprefix\endcsname\relax\def\urlprefix{URL }\fi
\providecommand{\bibinfo}[2]{#2}
\providecommand{\eprint}[2][]{\url{#2}}

\bibitem[{\citenamefont{Ring and Schuck}(1980)}]{Rin80}
\bibinfo{author}{\bibfnamefont{P.}~\bibnamefont{Ring}} \bibnamefont{and}
  \bibinfo{author}{\bibfnamefont{P.}~\bibnamefont{Schuck}},
  \emph{\bibinfo{title}{The Nuclear Many-Body Problem}}
  (\bibinfo{publisher}{Springer-Verlag, New York}, \bibinfo{year}{1980}).

\bibitem[{\citenamefont{Blaizot and Ripka}(1986)}]{Bla86}
\bibinfo{author}{\bibfnamefont{J.-P.} \bibnamefont{Blaizot}} \bibnamefont{and}
  \bibinfo{author}{\bibfnamefont{G.}~\bibnamefont{Ripka}},
  \emph{\bibinfo{title}{Quantum Theory of Finite Systems}}
  (\bibinfo{publisher}{MIT press, Cambridge}, \bibinfo{year}{1986}).

\bibitem[{\citenamefont{Tsoneva et~al.}(2004)\citenamefont{Tsoneva, Lenske, and
  Stoyanov}}]{Tso04}
\bibinfo{author}{\bibfnamefont{N.}~\bibnamefont{Tsoneva}},
  \bibinfo{author}{\bibfnamefont{H.}~\bibnamefont{Lenske}}, \bibnamefont{and}
  \bibinfo{author}{\bibfnamefont{C.}~\bibnamefont{Stoyanov}},
  \bibinfo{journal}{Phys. Lett. B} \textbf{\bibinfo{volume}{586}},
  \bibinfo{pages}{213} (\bibinfo{year}{2004}).

\bibitem[{\citenamefont{Sarchi et~al.}(2004)\citenamefont{Sarchi, Bortignon,
  and Col\`{o}}}]{Sar04}
\bibinfo{author}{\bibfnamefont{D.}~\bibnamefont{Sarchi}},
  \bibinfo{author}{\bibfnamefont{P.-F.} \bibnamefont{Bortignon}},
  \bibnamefont{and} \bibinfo{author}{\bibfnamefont{G.}~\bibnamefont{Col\`{o}}},
  \bibinfo{journal}{Phys. Lett. B} \textbf{\bibinfo{volume}{601}},
  \bibinfo{pages}{27} (\bibinfo{year}{2004}).

\bibitem[{\citenamefont{Petkov and Stoitsov}(1991)}]{Pet91}
\bibinfo{author}{\bibfnamefont{I.~Z.} \bibnamefont{Petkov}} \bibnamefont{and}
  \bibinfo{author}{\bibfnamefont{M.~V.} \bibnamefont{Stoitsov}},
  \emph{\bibinfo{title}{Nuclear Density Functional Theory}}
  (\bibinfo{publisher}{Clarendon press, Oxford}, \bibinfo{year}{1991}).

\bibitem[{\citenamefont{Fiolhais et~al.}(2003)\citenamefont{Fiolhais, Nogueira,
  and Marques}}]{Fio03}
\bibinfo{editor}{\bibfnamefont{C.}~\bibnamefont{Fiolhais}},
  \bibinfo{editor}{\bibfnamefont{F.}~\bibnamefont{Nogueira}}, \bibnamefont{and}
  \bibinfo{editor}{\bibfnamefont{M.}~\bibnamefont{Marques}}, eds.,
  \emph{\bibinfo{title}{A Primer in Density Functional Theory}}
  (\bibinfo{publisher}{Springer, Berlin}, \bibinfo{year}{2003}).

\bibitem[{\citenamefont{Arnould and Goriely}(2003)}]{Arn03}
\bibinfo{author}{\bibfnamefont{M.}~\bibnamefont{Arnould}} \bibnamefont{and}
  \bibinfo{author}{\bibfnamefont{S.}~\bibnamefont{Goriely}},
  \bibinfo{journal}{Phys. Rep.} \textbf{\bibinfo{volume}{384}},
  \bibinfo{pages}{1} (\bibinfo{year}{2003}).

\bibitem[{\citenamefont{Terasaki et~al.}(2005)\citenamefont{Terasaki, Engel,
  and \mbox{M. Bender et al.}}}]{Ter05}
\bibinfo{author}{\bibfnamefont{J.}~\bibnamefont{Terasaki}},
  \bibinfo{author}{\bibfnamefont{J.}~\bibnamefont{Engel}}, \bibnamefont{and}
  \bibinfo{author}{\bibnamefont{\mbox{M. Bender et al.}}},
  \bibinfo{journal}{Phys.\ Rev.\ C} \textbf{\bibinfo{volume}{71}},
  \bibinfo{pages}{034310} (\bibinfo{year}{2005}).

\bibitem[{\citenamefont{Yamagami and \mbox{Van Giai}}(2004)}]{Yam04}
\bibinfo{author}{\bibfnamefont{M.}~\bibnamefont{Yamagami}} \bibnamefont{and}
  \bibinfo{author}{\bibfnamefont{N.}~\bibnamefont{\mbox{Van Giai}}},
  \bibinfo{journal}{Phys. Rev. C} \textbf{\bibinfo{volume}{69}},
  \bibinfo{pages}{034301} (\bibinfo{year}{2004}).

\bibitem[{\citenamefont{\mbox{Pena Arteage} and Ring}(2008)}]{Pen08}
\bibinfo{author}{\bibfnamefont{D.}~\bibnamefont{\mbox{Pena Arteage}}}
  \bibnamefont{and} \bibinfo{author}{\bibfnamefont{P.}~\bibnamefont{Ring}},
  \bibinfo{journal}{Phys. Rev. C} \textbf{\bibinfo{volume}{77}},
  \bibinfo{pages}{034317} (\bibinfo{year}{2008}).

\bibitem[{\citenamefont{P\'{e}ru and Goutte}(2008)}]{Per08}
\bibinfo{author}{\bibfnamefont{S.}~\bibnamefont{P\'{e}ru}} \bibnamefont{and}
  \bibinfo{author}{\bibfnamefont{H.}~\bibnamefont{Goutte}},
  \bibinfo{journal}{Phys. Rev. C} \textbf{\bibinfo{volume}{77}},
  \bibinfo{pages}{044313} (\bibinfo{year}{2008}).

\bibitem[{\citenamefont{Yoshida and \mbox{Van Giai}}(2008)}]{Yos08}
\bibinfo{author}{\bibfnamefont{K.}~\bibnamefont{Yoshida}} \bibnamefont{and}
  \bibinfo{author}{\bibfnamefont{N.}~\bibnamefont{\mbox{Van Giai}}},
  \bibinfo{journal}{Phys. Rev. C} \textbf{\bibinfo{volume}{78}},
  \bibinfo{pages}{064316} (\bibinfo{year}{2008}).

\bibitem[{\citenamefont{\mbox{C. Losa et al.}}(2010)}]{Los10}
\bibinfo{author}{\bibnamefont{\mbox{C. Losa et al.}}} (\bibinfo{year}{2010}),
  \bibinfo{note}{\texttt{http://arxiv.org/abs/1002.4351v1}}.

\bibitem[{\citenamefont{Inakura et~al.}(2009)\citenamefont{Inakura,
  Nakatsukasa, and Yabana}}]{Ina09}
\bibinfo{author}{\bibfnamefont{T.}~\bibnamefont{Inakura}},
  \bibinfo{author}{\bibfnamefont{T.}~\bibnamefont{Nakatsukasa}},
  \bibnamefont{and} \bibinfo{author}{\bibfnamefont{K.}~\bibnamefont{Yabana}},
  \bibinfo{journal}{Phys. Rev. C} \textbf{\bibinfo{volume}{80}},
  \bibinfo{pages}{044301} (\bibinfo{year}{2009}).

\bibitem[{\citenamefont{\mbox{V. O. Nesterenko et al.}}(2006)}]{Nes06}
\bibinfo{author}{\bibnamefont{\mbox{V. O. Nesterenko et al.}}},
  \bibinfo{journal}{Phys. Rev. C} \textbf{\bibinfo{volume}{74}},
  \bibinfo{pages}{064306} (\bibinfo{year}{2006}).

\bibitem[{\citenamefont{\mbox{A. P. Severyukhin et al.}}(2008)}]{Sev08}
\bibinfo{author}{\bibnamefont{\mbox{A. P. Severyukhin et al.}}},
  \bibinfo{journal}{Phys. Rev. C} \textbf{\bibinfo{volume}{77}},
  \bibinfo{pages}{024322} (\bibinfo{year}{2008}).

\bibitem[{\citenamefont{\mbox{J. Toivanen et al.}}(2010)}]{Toi10}
\bibinfo{author}{\bibnamefont{\mbox{J. Toivanen et al.}}},
  \bibinfo{journal}{Phys. Rev. C} \textbf{\bibinfo{volume}{81}},
  \bibinfo{pages}{034312} (\bibinfo{year}{2010}).

\bibitem[{\citenamefont{Bohr and Mottelson}(1975)}]{Boh75}
\bibinfo{author}{\bibfnamefont{A.}~\bibnamefont{Bohr}} \bibnamefont{and}
  \bibinfo{author}{\bibfnamefont{B.~R.} \bibnamefont{Mottelson}},
  \emph{\bibinfo{title}{Nuclear Structure, vol. II}}
  (\bibinfo{publisher}{Benjamin, Reading}, \bibinfo{year}{1975}).

\bibitem[{\citenamefont{Soloviev and Shirikova}(1989)}]{Sol89}
\bibinfo{author}{\bibfnamefont{V.~G.} \bibnamefont{Soloviev}} \bibnamefont{and}
  \bibinfo{author}{\bibfnamefont{N.~Y.} \bibnamefont{Shirikova}},
  \bibinfo{journal}{Z. Phys.} \textbf{\bibinfo{volume}{A334}},
  \bibinfo{pages}{149} (\bibinfo{year}{1989}).

\bibitem[{\citenamefont{\mbox{M. V. Stoitsov et al.}}(2005)}]{Sto05}
\bibinfo{author}{\bibnamefont{\mbox{M. V. Stoitsov et al.}}},
  \bibinfo{journal}{Comp. Phys. Comm.} \textbf{\bibinfo{volume}{167}},
  \bibinfo{pages}{43} (\bibinfo{year}{2005}).

\bibitem[{\citenamefont{\mbox{A. Blazkiewicz et al.}}(2005)}]{Bla05}
\bibinfo{author}{\bibnamefont{\mbox{A. Blazkiewicz et al.}}},
  \bibinfo{journal}{Phys. Rev. C} \textbf{\bibinfo{volume}{71}},
  \bibinfo{pages}{054321} (\bibinfo{year}{2005}).

\bibitem[{\citenamefont{Dobaczewski et~al.}(1984)\citenamefont{Dobaczewski,
  Flocard, and J.Treiner}}]{Dob84}
\bibinfo{author}{\bibfnamefont{J.}~\bibnamefont{Dobaczewski}},
  \bibinfo{author}{\bibfnamefont{H.}~\bibnamefont{Flocard}}, \bibnamefont{and}
  \bibinfo{author}{\bibnamefont{J.Treiner}}, \bibinfo{journal}{Nucl.\ Phy.\ A}
  \textbf{\bibinfo{volume}{422}}, \bibinfo{pages}{103} (\bibinfo{year}{1984}).

\bibitem[{\citenamefont{\mbox{M. V. Stoitsov et al.}}(2003)}]{Sto03}
\bibinfo{author}{\bibnamefont{\mbox{M. V. Stoitsov et al.}}},
  \bibinfo{journal}{Phys. Rev. C} \textbf{\bibinfo{volume}{68}},
  \bibinfo{pages}{054312} (\bibinfo{year}{2003}).

\bibitem[{\citenamefont{Chabanat et~al.}(1998)\citenamefont{Chabanat, Bonche,
  Haensel, Meyer, and Schaeffer}}]{Cha98}
\bibinfo{author}{\bibfnamefont{E.}~\bibnamefont{Chabanat}},
  \bibinfo{author}{\bibfnamefont{P.}~\bibnamefont{Bonche}},
  \bibinfo{author}{\bibfnamefont{P.}~\bibnamefont{Haensel}},
  \bibinfo{author}{\bibfnamefont{J.}~\bibnamefont{Meyer}}, \bibnamefont{and}
  \bibinfo{author}{\bibfnamefont{R.}~\bibnamefont{Schaeffer}},
  \bibinfo{journal}{Nucl.~Phys.~A} \textbf{\bibinfo{volume}{635}},
  \bibinfo{pages}{231} (\bibinfo{year}{1998}).

\bibitem[{\citenamefont{Terasaki and Engel}(2006)}]{Ter06}
\bibinfo{author}{\bibfnamefont{J.}~\bibnamefont{Terasaki}} \bibnamefont{and}
  \bibinfo{author}{\bibfnamefont{J.}~\bibnamefont{Engel}},
  \bibinfo{journal}{Phys. Rev. C} \textbf{\bibinfo{volume}{74}},
  \bibinfo{pages}{044301} (\bibinfo{year}{2006}).

\bibitem[{\citenamefont{\mbox{J. Bartel et al.}}(1982)}]{Bar82}
\bibinfo{author}{\bibnamefont{\mbox{J. Bartel et al.}}},
  \bibinfo{journal}{Nucl. Phys. A} \textbf{\bibinfo{volume}{386}},
  \bibinfo{pages}{79} (\bibinfo{year}{1982}).

\bibitem[{\citenamefont{\mbox{C. Fahlander et al.}}(1992)}]{Fah92}
\bibinfo{author}{\bibnamefont{\mbox{C. Fahlander et al.}}},
  \bibinfo{journal}{Nucl. Phys. A} \textbf{\bibinfo{volume}{541}},
  \bibinfo{pages}{157} (\bibinfo{year}{1992}).

\bibitem[{\citenamefont{\texttt{http://www.nndc.bnl.gov}}()}]{natl_nucl_data}
\bibinfo{author}{\bibnamefont{\texttt{http://www.nndc.bnl.gov}}}.

\bibitem[{\citenamefont{Bohr and Mottelson}(1969)}]{Boh69}
\bibinfo{author}{\bibfnamefont{A.}~\bibnamefont{Bohr}} \bibnamefont{and}
  \bibinfo{author}{\bibfnamefont{B.~R.} \bibnamefont{Mottelson}},
  \emph{\bibinfo{title}{Nuclear Structure, vol. I}}
  (\bibinfo{publisher}{Benjamin, New York}, \bibinfo{year}{1969}).

\bibitem[{\citenamefont{Terasaki et~al.}(2008)\citenamefont{Terasaki, Engel,
  and Bertsch}}]{Ter08}
\bibinfo{author}{\bibfnamefont{J.}~\bibnamefont{Terasaki}},
  \bibinfo{author}{\bibfnamefont{J.}~\bibnamefont{Engel}}, \bibnamefont{and}
  \bibinfo{author}{\bibfnamefont{G.~F.} \bibnamefont{Bertsch}},
  \bibinfo{journal}{Phys. Rev. C} \textbf{\bibinfo{volume}{78}},
  \bibinfo{pages}{044311} (\bibinfo{year}{2008}).

\bibitem[{\citenamefont{\mbox{A. Voinov et al.}}(2001)}]{Voi01}
\bibinfo{author}{\bibnamefont{\mbox{A. Voinov et al.}}},
  \bibinfo{journal}{Phys. Rev. C} \textbf{\bibinfo{volume}{63}},
  \bibinfo{pages}{044313} (\bibinfo{year}{2001}).

\end{thebibliography}

\end{document}